\documentclass{svproc}
\usepackage{booktabs}       
\usepackage{tabulary}
\usepackage{graphicx}
\usepackage{float}
\usepackage{url}
\usepackage{hyperref}       

\begin{document}
\mainmatter              
\title{Towards Fairness in AI for Melanoma Detection: Systemic Review and Recommendations}
\titlerunning{AI for Melanoma}  
%
\author{Laura N Montoya \and Jennafer Shae Roberts \and
 Belén Sánchez Hidalgo}

 \authorrunning{Montoya et al.} 
 \tocauthor{Laura N Montoya, Jennafer Shae Roberts, and Belén Sánchez Hidalgo}

 \institute{Accel AI Institute, San Francisco, CA, USA,\\
 \email{laura@accel.ai} \\
 \email{jennafershae@accel.ai}\\
\email{belen.sanchez27@googlemail.com}
 }

\maketitle              

\begin{abstract}
Early and accurate melanoma detection is crucial for improving patient outcomes. Recent advancements in artificial intelligence (AI) have shown promise in this area, but the technology's effectiveness across diverse skin tones remains a critical challenge. This study conducts a systematic review and preliminary analysis of AI-based melanoma detection research published between 2013 and 2024, focusing on deep learning methodologies, datasets, and skin tone representation. Our findings indicate that while AI can enhance melanoma detection, there is a significant bias towards lighter skin tones. To address this, we propose including skin hue in addition to skin tone as represented by the L’Oreal Color Chart Map for a more comprehensive skin tone assessment technique. This research highlights the need for diverse datasets and robust evaluation metrics to develop AI models that are equitable and effective for all patients. By adopting best practices outlined in a PRISMA-Equity framework tailored for healthcare and melanoma detection, we can work towards reducing disparities in melanoma outcomes.
\keywords{Artificial Intelligence, Machine Learning, Deep Learning, Computer Vision, Melanoma, Skin Cancer, Dermatological imaging, Systemic Literature Review, Skin Tone, Hue}
\end{abstract}
\section{Introduction}
\begin{figure}[H]
  \centering
  \includegraphics[width=\linewidth]{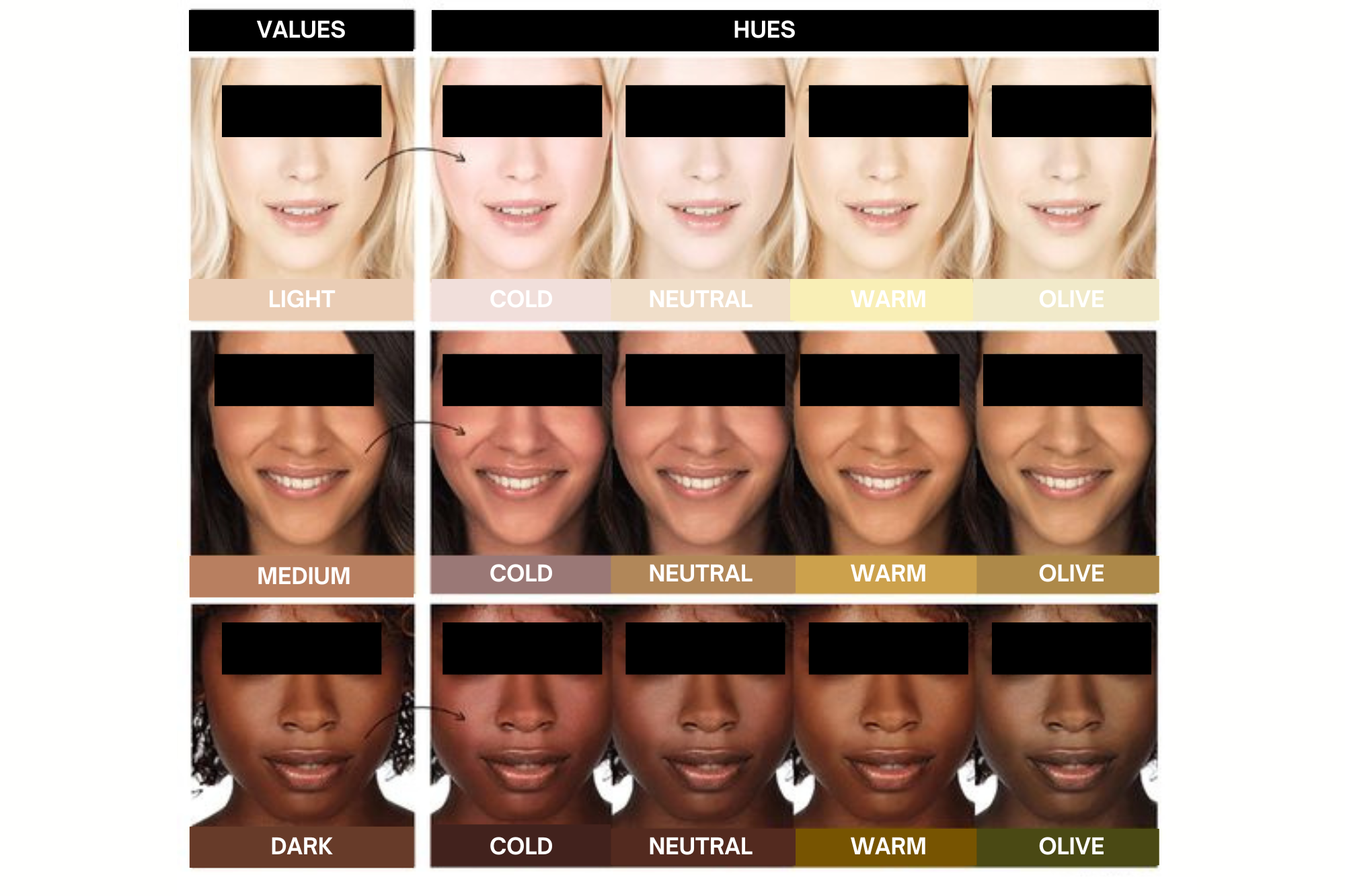}
  \caption{Depicts how hue undertone can affect skin color dimension. Image altered to protect model privacy and the text was translated from Portuguese to English. [Public domain], via Google Search for beauty palette demonstration. (\url{https://i.pinimg.com/564x/1a/5c/18/1a5c18940a453ac538708c5562834ff6.jpg}). Skin tone palette range with Values (light, medium, dark), and Hues (cold, neutral, warm, olive) depicted as a range of dimensionality.}
  \label{fig:skin-value-hue}
  \vspace{-4mm}
\end{figure}

\begin{figure}[H]
  \centering
  \includegraphics[width=\linewidth]{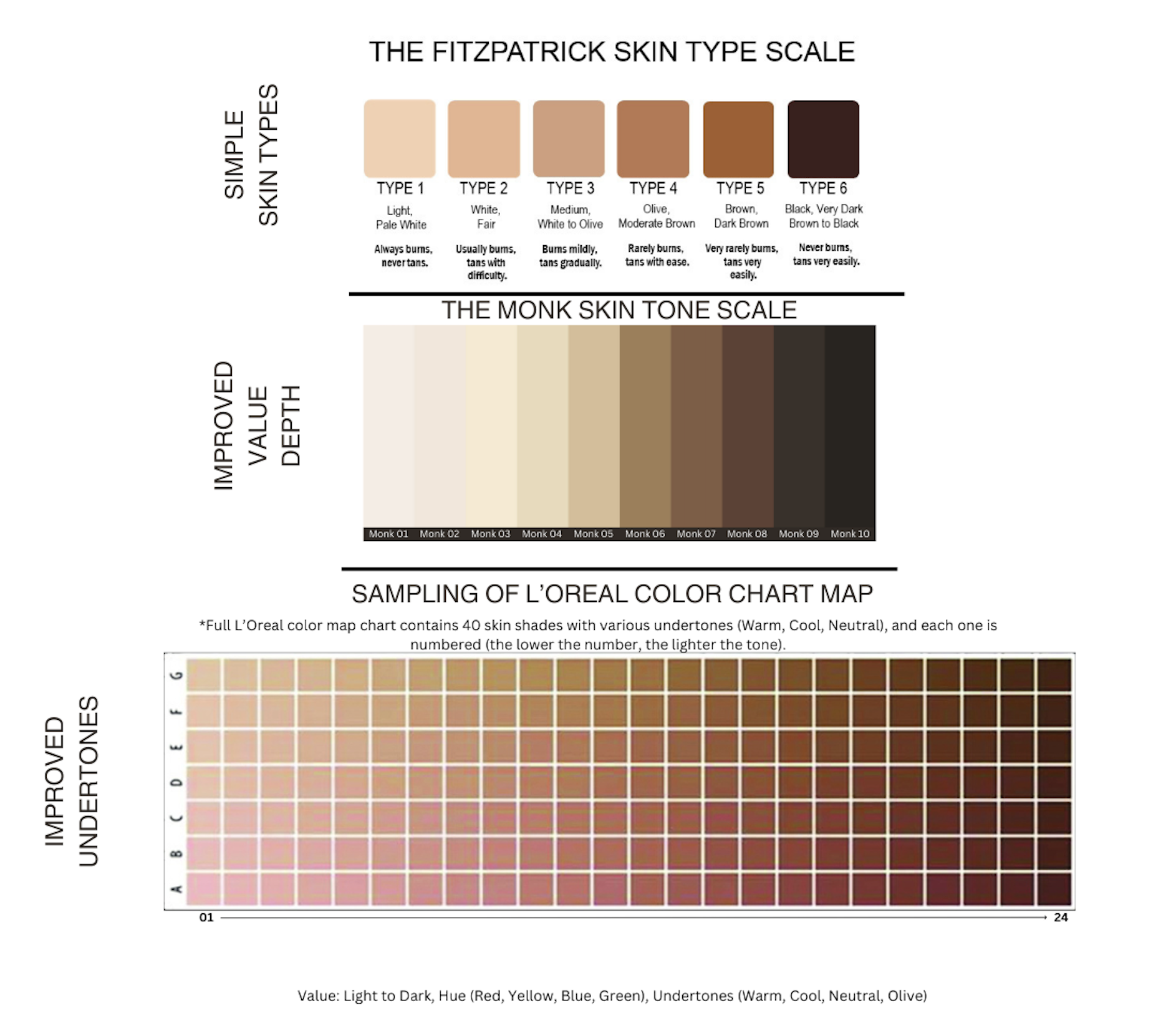}
  \caption{Comparison of skin tone scales that can be used for skin cancer detection utilizing AI. Recreation of Fitzpatrick Skin Type Scale, Monk Skin Tone Scale, and Sampling of L'Oreal Color Chart Map for reference.}
  \label{fig:skin-compare}
  \vspace{-6mm}
\end{figure}

Machine learning (ML) for early disease detection can save lives and is an excellent, ethical use case for this powerful technology. However, bias in healthcare as well as in datasets can cause disparities in who benefits, and who experiences marginalization, with deadly consequences. This paper highlights examples of how bias is perpetuated in image recognition algorithms for detecting melanoma, the deadliest form of skin cancer \cite{AmericanAcademyofDermatology2016PoorShows}. Algorithms used in dermatological applications are developed using images mostly from fair-skinned populations, neglecting darker skin types \cite{Guo2022BiasReview}. This can lead to unintentionally biased algorithms that perform poorly or unfairly on diverse patients. Since the majority of datasets of skin lesions contain only images of white skin \cite{Daneshjou2021DisparitiesImages}, this can lead to unintentionally biased algorithms that perform poorly or unfairly on diverse patients. This is critical, because although melanoma is much more common on lighter skin, the prognosis is statistically worse for dark-skinned patients \cite{KinyanjuiFairnessDermatology}.
 Early detection is key, as it greatly increases survival rates. ML Melanoma detection technology must be bias-free and accurate for people across all skin tones.  
ML algorithms reflect cognitive biases in humans, which can result in predictions and decisions that are “unfair”.
\cite{Angwin2022MachineBias}. Unfairness in ML decision-making equates to prejudice or favoritism based on inherent or acquired characteristics of an individual or group \cite{Mehrabi2021ALearning}. Within medical research, the term bias refers to “a feature of the design of a study, or the execution of a study, or the analysis of the data from a study, that makes evidence misleading” \cite{Stegenga2018CareMedicine} \cite{Pot2021NotRadiology}. If left unchecked, healthcare bias gets reproduced in ML through healthcare diagnosis and treatment at scale, which further disadvantages marginalized patients. Some biases can be easily detected and countered, such as through appropriate data curation; for example, having a balanced representation across skin tones and genders in training sets. However, in other cases, biases are hidden and untraceable \cite{Starke2021TowardsLearning}. In the case of detecting skin disease such as melanoma, skin tone is a factor that must be annotated correctly; however skin scales are quite limited, sometimes only split into two categories, light and dark values, if they are labeled at all. The most popular skin color scale currently being used for data annotation for image recognition techniques is the Fitzpatrick Skin Tone Scale (FST) \cite{Fitzpatrick1975SoleilPeau} which has 6 skin tones. Dating from the 1970s, it originally featured just 4 light tones and was designed for detecting photo sensitivity for white skin, with two darker tones added later \cite{Heldreth2023WhichIntelligence.}. The Monk Skin Scale was recently developed and still needs testing, but promisingly has 10 tones, 5 light and 5 dark \cite{Barrett2023SkinDatasets}. These scales can be viewed in figure \ref{fig:skin-compare}.
Recent research \cite{Thong2023BeyondColor} adds another axis, skin hue, which is described as ranging from red to yellow. This offers a more complete representation of variations of skin color by providing a multidimensional scale \cite{Thong2023BeyondColor}. Although the study ‘Beyond Skin Tone: A Multidimensional Measure of Apparent Skin Color’ introduces the use of hue in computer vision, it only covers red-yellow colors, and neglects a full-color palette which would include green-blue colors. Figure \ref{fig:skin-value-hue} demonstrates the range of dimensionality from skin tones. The effect of hue (blue, red, yellow, green) on skin tones adds depth to each face producing a range of undertones (cold, neutral, warm, and olive). In the realm of color theory, the concept of 'contrast of hue' emphasizes the distinctiveness among fundamental colors, with primary hues like yellow, red, and blue exhibiting the most pronounced differences \cite{Varner1984TemporalTheory}. Because skin cancer appears differently on different colored skin, it is important to acknowledge a full range of colors present in both healthy skin and suspicious lesions within datasets used to train skin cancer detection ML tools.

Although skin hues and undertones have been considered in skin color scales for cosmetology, as in the L’Oreal color match map \cite{2023FindSkin} which can be seen in figure \ref{fig:skin-compare}, skin color for dermatology datasets has been unidimensional, focusing only on tone. While color range in images has been shown to have a significant impact on the robustness of Deep Neural Networks \cite{Koenderink2015HueSpace}, this finding has yet to be adapted to skin cancer detection. These findings should correlate to AI for melanoma detection since the contrast between skin color and skin lesions is a preliminary marker during feature extraction. 
Although the  Fitzpatrick Skin Tone (FST) measurement scale is not diverse enough and leads to biased AI tools, it is continually used and has even been used to test a recently FDA-approved AI device for detecting melanoma. Health technology company DermaSensor Inc. has shared encouraging findings from a sub-analysis of the DERM-SUCCESS study \cite{BusinessWire2023DermaSensorPatients}, showcasing the effectiveness of their handheld Elastic Scattering Spectroscopy (ESS) device in detecting skin cancer. The analysis specifically examined the device's performance in patients with all 6 Fitzpatrick skin types \cite{Bradford2009AcralMelanoma} \cite{BusinessWire2023DermaSensorPatients}. However, upon closer examination of the study, only 1.8 percent of the 1,005 patients represented in the study were on FST VI, the darkest skin tone on the scale, and 97.1 percent were reported as white \cite{Merry2023ClinicalObjective}. This raises concerns about the potential inaccuracy of this FDA-approved AI melanoma detection device for people with dark skin, who are disproportionately affected by fatal melanoma, and yet underrepresented in testing.

Our goal is to establish an ethical AI framework for melanoma detection that prioritizes diverse skin tone representation. We advocate for the adoption of improved scales like the Monk and L'Oreal maps. Future studies should ensure equitable representation and testing across skin tones to guarantee AI's effectiveness for all. Please refer to tables 2 through 7 in the discussion section for further recommendations for curating a diverse dataset, including purpose, ownership, funding, and data annotation, as well as recommendations for each stage of the data life cycle. 

We begin this systemic review by discussing the findings and limitations of related works. Section 2 outlines the materials and methods employed in this study. We detail our objectives, eligibility criteria, information sources and search strategy, data collection and analysis procedures, and acknowledge potential limitations. Section 3 presents our preliminary analysis of existing literature, image datasets for melanoma detection, and skin tone scales. In Section 4, we discuss our findings and propose recommendation tables for more equitable machine learning practices, particularly in medical applications and skin cancer detection. Finally, Section 5 concludes our study by summarizing key findings and implications of this systemic literature review for future research.

    \subsection{Related Works Supporting Rationale}

    Our research began with reviewing literature reviews published between 2021 and 2024 on 'AI for Melanoma Detection' to identify their focus and limitations. Below are summaries of the papers which were the most relevant to our research. Each of them could benefit from expanding the skin color scales used to include a wider range of skin tones as well as skin hues. 

\subsubsection{Assessing the Generalizability of Deep Neural
Networks-Based Models for Black Skin Lesions \cite{Barros2023AssessingLesions}}

Observing that previous research on AI-based melanoma detection primarily focuses on light-skinned populations, this study addresses this gap by evaluating supervised and self-supervised models on skin lesion images extracted from acral regions commonly observed in Black individuals. The authors carefully curated a dataset of acral skin lesions and assessed model performance using the Fitzpatrick scale. Their results highlight the poor generalizability of existing models to darker skin tones, emphasizing the urgent need for diverse datasets and specialized models to ensure equitable access to AI-powered melanoma diagnosis. However, the study did not mention skin hue as a part of skin color, and is limited by the use of the Fitzpatrick scale and not a more representative skin color scale. 

\subsubsection{Fairness of Classifiers Across Skin Tones in Dermatology \cite{KinyanjuiFairnessDermatology}}

This study investigates the representation of skin tones in dermatological image datasets and its potential impact on model performance. The authors estimated skin tone using the Individual Typology Angle (ITA) method on two widely used datasets. They found that the majority of images represent lighter skin tones, suggesting the underrepresentation of darker skin individuals. While there was no clear correlation between model accuracy and skin tone in this analysis, the study highlights the need for more diverse datasets to ensure fairness and effectiveness in skin disease diagnosis using AI. However, the study has several limitations. First, the ITA method, while commonly used, may not accurately capture the full spectrum of skin tones, particularly for darker skin tones. Second, the analysis focused on two widely used datasets, which may not represent the entire diversity of dermatological images available. Finally, the study did not explore the potential impact of other factors, such as image quality, lighting conditions, and the presence of artifacts, on model performance.

\subsubsection{Revisiting Skin Tone Fairness in Dermatological
Lesion Classification \cite{Kalb2023RevisitingClassification}}

This paper investigates the challenges of ensuring fairness in skin cancer classification models. Due to the lack of skin tone labels in public datasets, and as in the previously mentioned study, researchers often estimate skin tone using the Individual Typology Angle (ITA) method. However, this study finds significant inconsistencies between different ITA-based approaches. The authors attribute this to limitations in the dataset used for evaluation. They conclude that more robust skin tone estimation methods and diverse datasets with accurate skin tone labels are essential for developing fair AI models in dermatology. The study also highlights the limitations of relying solely on ITA for skin tone estimation, as it may not accurately capture the full spectrum of skin tones, particularly for darker skin tones. Additionally, the lack of diverse datasets with accurate skin tone labels remains a significant challenge in developing fair AI models in dermatology.

    \subsubsection{Skin Deep: Investigating Subjectivity in Skin Tone Annotations for Computer Vision Benchmark Datasets \cite{Barrett2023SkinDatasets}}
    This study found that while using skin tone instead of race for fairness evaluations in computer vision seems objective, the annotation process remains biased by human annotators. Untested scales, unclear procedures, and a lack of awareness about annotator backgrounds and social context significantly influence skin tone labeling. This study exposes how even minor design choices in the annotation process, like scale order (dark to light instead of light to dark) or image context (face or no face, skin lesion presence), can sway agreement and introduce uncertainty in skin tone assessments. The researchers emphasize the need for greater transparency and a critical lens on subjectivity to ensure fairer and more robust evaluations in computer vision. This study highlights the challenges of ensuring objectivity and consistency in skin tone annotations, even when using seemingly objective methods. The researchers emphasize the need for greater transparency, standardized procedures, and careful consideration of annotator biases to mitigate these challenges and ensure fairer and more robust evaluations in computer vision.

    \subsubsection{Out-of-Distribution Detection in Dermatology using Input Perturbation and Subset Scanning \cite{Kim2021Out-of-DistributionScanning}}
    This paper emphasizes the importance of detecting outliers in testing that differ significantly from training data to reduce biased or unfair diagnoses. Identifying out-of-distribution (OOD) samples is crucial before making decisions to ensure that knowledge transfer from in-distribution training samples to OOD test samples is principled and extendable to novel scenarios. Furthermore, DL solutions, including OOD detectors, must ensure equivalent detection capability across sub-populations. In dermatology, insufficient representation of dark skin tones in academic materials and clinical care is a growing concern that contributes to disparities in patient care and worsens with the use of AI machine learning algorithms trained on imbalanced datasets primarily consisting of light skin tones. The paper demonstrates that these datasets disproportionately under-represent darker skin tones. 

    \subsubsection{Lack of Transparency and Potential Bias in Artificial Intelligence Data Sets and Algorithms A Scoping Review \cite{Daneshjou2021LackAlgorithms}}
   This scoping review found that dermatology AI studies often lack detailed dataset descriptions, use inconsistent disease labels, and lack transparency in reporting patient demographics. These issues hinder reproducibility, accuracy, and fairness in AI models \cite{Daneshjou2021LackAlgorithms}. The authors addressed this with the Diverse Dermatology Images (DDI) dataset, curated in 2022. The authors stated that there were no public AI benchmarks that contained images of biopsy-proven malignancy on dark skin before the DDI dataset \cite{Daneshjou2021DisparitiesImages}. The images in the DDI dataset include a sampling across FST I-VI skin tones for direct comparison of testing on images labeled with the lightest FST I/II (control group) and darkest FST V/VI skin-tone categories. Training and testing by the Stanford researchers is described in ‘Disparities in Dermatology AI: Assessments Using Diverse Clinical Images’ \cite{Daneshjou2021DisparitiesImages}. Future enhancements include the recommendation to retest this dataset against the Monk skin tone scale, as it is more inclusive of darker skin tones, as well as considering skin hue for red and blue pigmentation \cite{Thong2023BeyondColor}.

\section{Materials and Methodology}
    \subsection{Objectives}
    The primary objective of this systemic literature review is to evaluate the effectiveness and accuracy of AI models in detecting melanoma skin cancer. We aim to develop a framework for ethical melanoma detection utilizing AI models from our findings which reduces bias and accounts for the widest diversity in datasets for fair application. In our systematic review on AI for melanoma detection, we employed the PRISMA (Preferred Reporting Items for Systematic Reviews and Meta-Analyses) \cite{Page2021TheReviews} guidelines due to their emphasis on standardization, transparency, and comprehensive literature aggregation. PRISMA's structured approach ensures a thorough and unbiased collection of studies, critical for the dynamic field of AI. It aids in rigorous quality assessment, enhances comparability with similar studies, and ensures clear reporting. This methodological rigor aligns with our commitment to producing a review that is methodologically sound, ethically rigorous, and globally recognized, thereby enhancing the credibility and utility of our research in this evolving field. We followed the PRISMA-Equity reporting guidelines which is the standard for systemic literature reviews in medical research. Since the focus of our paper is AI for Melanoma detection, this research falls within the scope of medical research while improving upon computer vision research and it's capabilities. It has become more common for AI researchers to borrow from medical research standards in areas of privacy, equity, and transparency, and the use of the PRISMA guidelines in this context extends that practice demonstrating continued improvements for the AI research community towards protection for end-users of the technology.

    \subsection {Eligibility Criteria}
    We considered many criteria for including articles in this systematic review and preliminary analysis. The inclusion criteria consisted of papers published within the last 11 years between 2013 and 2024 in English, open-access with PDFs available online, utilized AI or ML models for melanoma detection, and peer-reviewed. We excluded papers that were not published in English, were closed access or behind a paywall, and that utilized models for the detection of other skin diseases non-inclusive of melanoma.

    \subsection {Information Sources and Search Strategy}
    Our search strategy included utilizing many common databases for scientific research including Google Scholar, PubMed, IEEE Xplore, and ScienceDirect. We also used AI or ML-based discovery tools including Litmaps and Elicit to find related research and extract relevant information for our preliminary review and meta-analysis. We searched terms relating to artificial intelligence (e.g., “AI”, “machine learning”, “deep learning”) combined with terms related to melanoma detection (e.g., “melanoma detection”, “skin cancer diagnosis”, “dermatological imaging”) as well as skin tone scales (“Monk Skin Tone Scale” and “Fitzpatrick Skin Tone Scale”). Search strings were created combining the various terms with Boolean operators. For example, ("AI" OR "machine learning" OR "deep learning") AND ("melanoma detection" OR "skin cancer diagnosis" OR "dermatological imaging"). Finally, we found related research in the references of other research papers. We filtered for studies published between January 2013 and December 2024. We excluded papers that were not open-access or did not have a PDF available online to review.  

    For Google Scholar, we made use of the SerpAPI Python extension to scrape results and export to a JSON object. Then we wrote the JSON to a CSV file for additional analysis.

    \subsection {Data Collection and Analysis}
    The articles found were listed in a table and split evenly among three reviewers who independently screened titles, abstracts, and content for eligibility. Full-text articles with PDFs available under open access were reviewed for manual and automated data extraction. Discrepancies were resolved through discussion or consultation with the other reviewers for missing or unclear information.

    Extracted data included the title, abstract, author(s), year of publication, and keywords for reference. To compare the datasets used in each study for quality and diversity of data, we gathered details on the image dataset size (number of images), whether the dataset was novel, and whether information on demographic diversity (age, gender, race, or ethnicity of patients), clinical diversity (skin type, lesion type, anatomical location of lesion), or image characteristics (source, imaging techniques, resolution, and whether the images were real or artificially generated) were included in the publication. To analyze the experimental design, we recorded whether there was any preprocessing or analysis of the data before training and testing, the type of AI model used, benchmarking comparison methods, outcomes (e.g., sensitivity, specificity, accuracy), and limitations. 

    To track ethical considerations on the use of AI models for melanoma detection, we noted whether the studies accounted for differences in skin tone through skin-tone analysis or skin-tone scales and whether they discussed issues related to privacy, accountability, or transparency of their models.

    Trends in the included literature review for 'publications over time', and 'citations over time', were identified and visualized through the use of the Python scientific libraries PANDAS for data analysis and manipulation, Matplotlib for visualizations, and Seaborn for statistical data visualization. 
    
     \subsection{   Limitations}
        While this systemic review provides a comprehensive review of the literature on fairness in AI for melanoma detection, it is primarily based on existing research. To validate the proposed recommendations or frameworks, continuing work is necessary to complete empirical analysis and experiments. Additionally, the suggested adoption of new skin tone scales, while beneficial, may face practical challenges in implementation. Furthermore, while the paper strongly advocates for specific skin tone scales, it's important to note that other methods or tools might also effectively address fairness issues in AI for melanoma detection. Finally, while the study addresses fairness in AI, it could benefit from further exploration of the practical implementation of these recommendations in real-world clinical settings. Potential obstacles and the feasibility of widespread adoption should be considered to ensure that the proposed solutions are not only theoretically sound but also practically viable.

\section{Results}
In our comprehensive review, we found a total of 665 studies, out of which 270 were accessible in full-text PDF format. The findings from our analysis reveal a discernible trend towards enhanced accuracy in melanoma detection when AI is utilized in conjunction with skin tone analysis. Notably, the incorporation of skin tone scales in AI methodologies shows substantial potential in improving melanoma detection, particularly in reducing biased outcomes. This integration appears to be a pivotal factor in the effectiveness and reliability of AI-driven diagnostics in dermatology.
    \subsection{Preliminary Analysis of Literature}
\begin{figure}[H]
        \centering
        \includegraphics[width=0.7\textwidth]{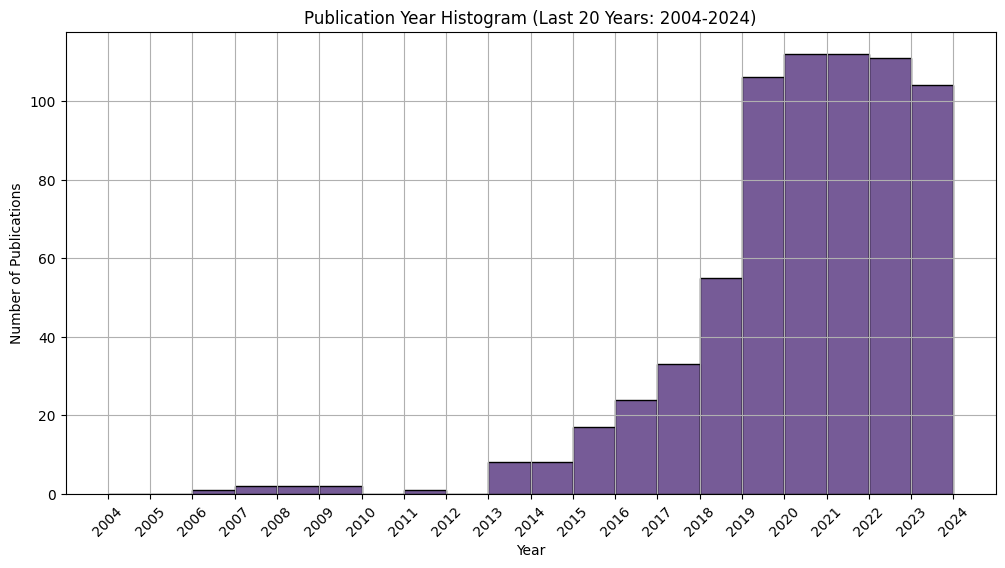}
        \caption{Number of publications related to AI and melanoma published during the last 20 years. The x-axis represents the years, and the y-axis represents the number of publications per year. The bins of the histogram correspond to individual years.  }
        \label{fig:histo-years}
        \vspace{-7mm}
    \end{figure}

    The publication year histogram, Figure \ref{fig:histo-years} provides a visual representation of the distribution of research papers in our literature review over the last 20 years.  We can notice a growing research interest and activity over the last two decades, with particular years standing out due to a higher volume of publications.

    From this graph, we can see a general increase in the number of publications over time.  The early years, specifically from 2003 to 2013, show a relatively lower number of publications that could suggest a nascent stage of research activity in this field. A notable increase in publications began in 2014, with a significant upward trend that continues until 2022.  After this year, there is a slight drop in the number of publications in 2023.  

    \begin{figure}[H]
        \centering
        \includegraphics[width=0.7\textwidth]{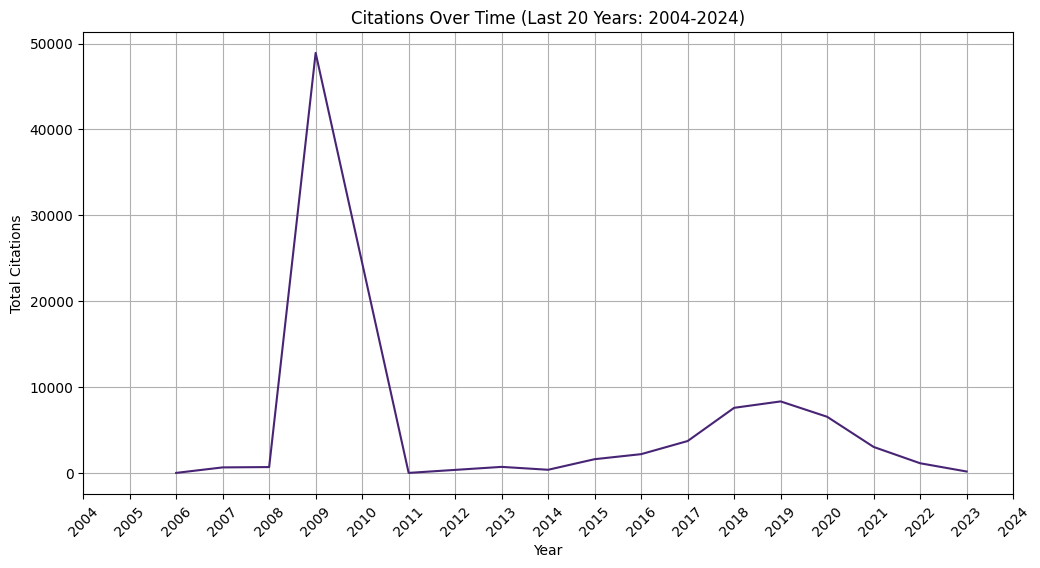}
        \caption{Number of citations related to publications about AI and melanoma over time from 2004 to 2023.  The x-axis denotes the years, while the y-axis indicates the total number of citations for that year. }
        \label{fig:num-cite}
    \end{figure}

    We found similar patterns in citations of AI melanoma research as the publications over time. As shown in Figure \ref{fig:num-cite}, following the peak in citations in 2009, there is a steep decline in citations over the next few years, and subsequently, there is a gradual increase from 2016, leading to a moderate peak in 2019.  From there, there is a decline followed by a plateau.  In sum, this provides an insightful view into the flow of research impact in the field of AI and melanoma detection, with significant fluctuations over time.

\subsection{Image Datasets for Melanoma Detection}
We compared the datasets used in the studies that were included in our literature review, as depicted in Table \ref{tab:datasets-compare}. The datasets vary significantly in size, from smaller datasets like PH2 with 200 images to larger ones like ASAN with 120,780 images. Larger datasets can offer more comprehensive training data, which may improve the model’s performance and generalizability. There is variability in the type and extent of meta-data provided. For example, some datasets include detailed clinical annotations and demographic information, while others offer limited or unspecified meta-data. Comprehensive meta-data helps in developing models that are more robust and fair across different population groups. Datasets like PAD-UFES-20 and Diverse Dermatology Images (DDI) provide detailed information on skin types and tones, which is crucial for developing models that perform well across diverse skin types. This is particularly important for minimizing biases and ensuring equitable detection performance. Of the datasets identified in our review, only 3 include skin tone identification, all of which utilize the Fitzpatrick Skin Type Scale. Only 7 of the 18 datasets included ethnicity in the provided meta-data.

Our literature review revealed a significant reliance on the Fitzpatrick Skin Type Scale in existing AI models for melanoma detection published in the last three years. While the Fitzpatrick Scale, developed in the 1970s, has been a longstanding tool in dermatology, our analysis highlights its limitations, particularly in terms of its simplistic categorization which does not fully encompass the diversity of global skin tones. 

In contrast, Figure \ref{fig:skin-compare} demonstrates the more recent Monk Skin Tone Scale offers a broader and more nuanced categorization of depth. This scale demonstrates an enhanced ability to differentiate subtle variances in skin tones through light and dark values, which is crucial for the accuracy of AI algorithms in detecting melanoma across diverse populations. However, despite its advancements, the Monk Scale is still limited in its representation of the full spectrum of skin tones, especially in underrepresented groups. Our recommendation gravitates towards the L'Oréal Color Chart Map, an innovative approach that significantly expands on the existing scales with the addition of undertones (cold, neutral, warm, olive) from Hue (yellow, red, blue, green).


\begin{table}[H]
    \centering
    \resizebox{\columnwidth}{!}{%
    \begin{tabular}{p{0.25\linewidth} | p{0.05\linewidth}|p{0.3\linewidth} |p{0.45\linewidth} |p{0.05\linewidth} }
    \textbf{Dataset Name} & \multicolumn{1}{l}{\textbf{Number of Images}} & \textbf{Skin Types and Skin Tones} & \textbf{Meta-data} & \multicolumn{1}{l}{\textbf{Year Published}} \\ \hline
    
    PH2 dataset & 200 & NS* & Medical annotation of all images (medical segmentation of the lesion, clinical and histological diagnosis, and the assessment of several dermoscopic criteria) (colors; pigment network; dots/globules; streaks; regression areas; blue-whitish veil) & 2012 \\ \hline
    
    Edinburgh Dermofit Library & 1,300 & NS* & Gender, Age, Ethnicity, binary segmentation mask & 2013 \\ \hline
    
    MED-NODE & 170 & NS* & NS* & 2015 \\ \hline
    
    Dermnet & 23,000 & NS* & Ethnicity & 2016 \\ \hline
    
    ISIC\_2016 & 1,279 & NS* & NS* & 2016 \\ \hline
    
    ISIC\_2017 & 2,600 & NS* & NS* & 2017 \\ \hline
    
    ASAN & 120,780 & NS* & Gender, Age, Ethnicity & 2018 \\ \hline
    
    HAM10000 dataset & 10,015 & NS* & Gender, Age & 2018 \\ \hline
    
    ISIC\_2018 & 11,527 & NS* & NS* & 2018 \\ \hline
    
    ISIC\_2019 (BCN\_20000) & 33,569 & NS* & NS* & 2019 \\ \hline
    
    ISIC\_2020 & 44,108 & NS* & ID of the patient, ID of the lesion, gender, age, anatomical site, and diagnosis & 2020 \\ \hline
    
    Normal & 48,271 & NS* & Gender, Age, Ethnicity & 2020 \\ \hline
    
    PAD-UFES-20 & 2,299 & Fitzpatrick skin type & age, skin lesion location, Fitzpatrick skin type, and skin lesion diameter & 2020 \\ \hline
    
    SNU & 2,201 & NS* & Gender, Age, Ethnicity & 2020 \\ \hline
    
    Web & 51,459 & NS* & Gender, Age, Ethnicity & 2020 \\ \hline
    
    Diverse Dermatology Images (DDI) dataset & 656 & 208 images of FST I–II (159 benign and 49 malignant), 241 images of FST III–IV (167 benign and 74 malignant), 207 images of FST V–VI (159 benign and 48 malignant) & Gender, Age, Ethnicity, Skin Tone & 2021 \\ \hline
    
    SIIM-ISIC Melanoma & 33,126 & NS* & Age & 2021 \\ \hline 
    
    SkinCon & 3,230 & Fitzpatrick skin type & NS* & 2023\\ \hline
    \end{tabular}%
    }
    \caption{Comparison of datasets used in our AI for Melanoma Detection Systemic Literature Review. Included in this table are the dataset name, number of images, whether skin tone was analyzed, meta-data types, and year of publication. *NS - Not Specified in Dataset description.}
    \label{tab:datasets-compare}
    \vspace{-8mm}
\end{table}

\subsection{Skin Tone Scales}

\section{Discussion}

There is a historically inadequate representation of darker skin tones in dermatology image datasets. The first iteration of the Fitzpatrick Skin Type (FST) \cite{Fitzpatrick1975SoleilPeau} scale was primarily designed for Caucasians, thereby disregarding other ethnic types, such as Black or Asian skin tones. Despite subsequent updates to these measures, they still failed to address the representation of darker skin tones \cite{Ware2020SKINType} serving as evidence of this neglect \cite{Arosarena2015OptionsPhototypes} \cite{Lester2020AbsenceManifestations}. While computer vision (CV) projects like Gender Shades\cite{Buolamwini2018Gender} utilized skin tone annotations to analyze facial recognition technology, limitations remain. Newer scales like Google's Monk Skin Tone Scale\cite{Monk2022DevelopingScale} aim to represent a wider range but lack thorough evaluation. Additionally, understanding the complex social context of skin tone is crucial for accurate annotation \cite{Barrett2023SkinDatasets}.

To accurately detect melanoma, diverse datasets representing the global range of skin tones and disease presentations are essential. Images must showcase a wide range of ethnicities, ages, genders, and skin types, with adequate representation of outliers \cite{Barrett2023SkinDatasets}. Accurate image annotation demands a meticulous approach. Expert dermatologists from diverse backgrounds must carefully categorize lesions, going beyond the Fitzpatrick scale to encompass the full range of skin colors (with undertones of red, yellow, blue, green) to ensure representation of all skin types. \cite{Thong2023BeyondColor}. While progress is being made, building truly inclusive datasets requires further research and collaboration to capture the vast tapestry of human skin color. The collection and utilization of diverse skin tone datasets must adhere to stringent ethical guidelines. These data should be anonymized and protected to prevent misuse. Robust data governance and security protocols are essential to safeguard patient privacy and prevent any potential discriminatory applications.

Future research should prioritize the development of more comprehensive skin tone scales to accurately capture the diversity of human skin. While the L'Oreal Color Chart Map offers a promising foundation, further refinement and validation are necessary to maximize its effectiveness in dermatological applications. Empirical studies investigating the impact of expanded skin tone scales on model performance are warranted to solidify the benefits of this approach.

We have depicted our recommendations for curating a diverse dataset in table(s) \ref{tab:dataset-recommendations}, \ref{tab:dataset-recommendations-2}, \ref{tab:dataset-recommendations-3}, and our recommendations for inclusive model development in tables \ref{tab:model-recommendation}, \ref{tab:model-recommendation-3}, \ref{tab:model-recommendation-5}. The tables are inspired by the PRISMA-Equity reporting guidelines \cite{Welch2012PRISMA-EquityEquity} and include suggestions from the Mitigating Bias in Machine Learning book chapter 9 focused on AI healthcare bias \cite{Roberts2024MitigatingLearning}, the Datasheets for Datasets \cite{Gebru2021DatasheetsDatasets}, and Model Cards for Model Reporting \cite{Mitchell2018ModelReporting} articles. 

The book chapter 'Towards Rectification of Machine Learning Bias in Healthcare Diagnostics' \cite{Roberts2024MitigatingLearning} presents a case study of skin cancer detection across diverse ethnic groups. It discusses bias types and mitigation strategies for each phase of the machine learning lifecycle. Datasheets for Datasets propose a standardized method for documenting datasets, addressing the lack of consistent documentation. Model Cards for Model Reporting introduce a framework for documenting the performance characteristics of trained machine learning models. We adapted these recommendations for standard image curation, healthcare settings, and AI melanoma detection.

The tables provide a comprehensive framework for developing and utilizing image datasets in AI-driven melanoma detection. They emphasize the importance of ethical considerations, data diversity, robust model development, and transparent reporting. By aligning the tables with the PRISMA-Equity guidelines, we aim to ensure a focus on fairness and inclusivity throughout the research process. By adhering to these guidelines, researchers can develop AI models that are more equitable and effective in detecting melanoma across diverse populations.

Included are recommendations for standard image curation with an ethical focus, extended recommendations specific to healthcare applications, and further extended recommendations for melanoma detection, reflective of the case study in this paper. This serves as a practical guide as well as an example to showcase the utility of the recommendations that are presented. Tables 2-7 outline a structured approach to developing and utilizing image datasets for AI-driven melanoma detection. These tables emphasize the importance of ethical considerations, data diversity, and robust model development throughout the research process. 

Table 2 highlights the need for diverse teams and clear ownership, while Table 3 underscores the significance of ethical data collection. Tables 4 and 5 delve into data annotation, preprocessing, and analysis, emphasizing the importance of skin tone representation. The subsequent tables focus on model development, evaluation, and reporting, with a strong emphasis on fairness and transparency. 

\section{Conclusion}

We recommend that the training and testing of AI models for skin cancer detection should explicitly incorporate a diverse range of skin tones and hues. This approach recognizes the significant variance in skin color across the human population, analogous to the primary, secondary, and tertiary contrasts in color theory. By doing so, the AI models can be more accurately tuned to detect melanoma across a broader spectrum of skin types, enhancing both the sensitivity and inclusivity of these diagnostic tools. Our comprehensive survey of 665 studies, including 270 full-text PDFs, underscores a pivotal advancement in melanoma detection accuracy through the integration of AI with skin tone analysis. The results from our systemic review highlight the significant potential of AI in the field of dermatology, especially when augmented by skin tone scales. The convergence of AI and skin tone considerations marks a promising direction in dermatological diagnostics, offering a more inclusive and effective framework for early melanoma detection. Our literature review underscores the limitations of the widely-used Fitzpatrick Skin Type Scale in AI melanoma detection, particularly its inadequate representation of global skin tone diversity. While the Monk Skin Tone Scale offers improved granularity, it still falls short in encompassing the full spectrum of skin tones. Our recommendation strongly favors the adoption of the L'Oréal Color Chart Map, which provides a more extensive and precise categorization of skin tones. This comprehensive scale not only enhances the accuracy and inclusivity of AI models in detecting melanoma but also aligns with the evolving needs of personalized healthcare. The L'Oréal Chart's diverse and data-driven approach significantly addresses the underrepresentation of certain skin tones in dermatological AI applications, making it a superior choice for future research and development in the field.

\begin{table}[H]
    \centering
    \resizebox{\columnwidth}{!}{%
    \begin{tabular}{p{0.15\linewidth} | p{0.06\linewidth}|p{0.3\linewidth} |p{0.3\linewidth} |p{0.3\linewidth} }
        \textbf{Category} &  \textbf{Item} & \textbf{Standard Image Curation with Ethical Focus} & \textbf{Extension for Healthcare} & \textbf{Extension for Melanoma Detection} \\ \hline
        \textbf{Purpose, ownership, and funding} & & & & \\ \hline
        & 1 & Define the specific purpose and/or recommended use case for the dataset. & Describe illness or ailment being diagnosed or treated through the use of the dataset for model training. & Describe the type of melanoma or skin lesion, the purpose of the dataset for robustness, the type of model, and the end application. \\ \hline
        & 2 & Specify the team, research group that will create the dataset and on behalf of which entity. & This team should include healthcare professionals (e.g., dermatologists, oncologists), data scientists, and ethicists to ensure comprehensive expertise. Specify the healthcare institution, research organization, or collaborative partnership leading the effort. Include details about their qualifications, previous experience in similar projects, and their role in ensuring the dataset's adherence to ethical and clinical standards. & Team: Describe the specific expertise required, including dermatologists with expertise in melanoma, medical researchers, and machine learning specialists. Emphasize the inclusion of specialists experienced in skin imaging and melanoma detection.  Entity: Specify the entity overseeing the dataset creation, such as a specialized melanoma research center, a university department with a focus on dermatology and artificial intelligence, or a consortium of healthcare institutions. Highlight any affiliations with reputable melanoma research initiatives or clinical practice guidelines. Governance: Include the governance structure for the dataset creation process including any review boards or committees overseeing the project to ensure it meets both healthcare and research standards. \\ \hline
        & 3 & Indicate who is funding the creation of the dataset. (Associated grant, grantor). If this is relevant, check for any funding biases. &  Specify the entities funding the dataset creation, including grants from organizations like research foundations, medical institutions, or industry partners specializing in treatment (dermatology and oncology). Detail the associated grant numbers, grantors, and their roles in the project. & Consider whether the funders have any specific interests in the outcomes of the melanoma detection research, which could affect dataset design, model development, or interpretation of results. \\ \hline
 \end{tabular} %
    }
    \caption{PRISMA-Equity: Recommendations for curating a diverse dataset - Purpose, ownership, and funding}
    \label{tab:dataset-recommendations}
    \vspace{-4mm}
\end{table}

\begin{table}[H]
    \centering
    \resizebox{\columnwidth}{!}{%
     \begin{tabular}{p{0.15\linewidth} | p{0.06\linewidth}|p{0.3\linewidth} |p{0.3\linewidth} |p{0.3\linewidth} }
        \textbf{Category} & \textbf{Item} & \textbf{Standard Image Curation with Ethical Focus} & \textbf{Extension for Healthcare} & \textbf{Extension for Melanoma Detection} \\ \hline
        \textbf{Data Collection} & & & &\\ \hline 
        & 4 & Define strategies for global data collection to ensure geographic diversity and plan to collect features / markers needed for bias testing. & Include a wide range of ethnicities, ages, genders, and skin types. & Collect a clinically diverse set of melanoma images including differing sizes, colors, shapes, stages of growth, both benign and malignant. \\ \hline
        & 5 & Note whether any data was excluded and why. & S* & S* \\ \hline
        & 6 & Specify the limitations of the dataset. & S* & S* \\ \hline
        & 7 & Adhere to ethical guidelines in data collection and use & Check if the dataset contains information that might be considered confidential or protected by legal privilege or doctor patient confidentiality. If it does, provide a description. & S* \\ \hline
        & 8 & Comply with data privacy best practices and regulations over the use of personal data. Consider international, national and local regulations and frameworks. Eg. GDPR in the European Union, PIPEDA in Canada, CCPA in California - USA. & Comply with data privacy best practices and regulations over the use of personal data. Consider international, national and local regulations and frameworks. Eg. HIPAA in USA, Declaration of Helsinki with ethical principles for medical research. & Implement stringent measures to anonymize and secure skin images and related medical information to prevent unauthorized access and maintain confidentiality. Describe the technical and administrative safeguards used to protect patient data. \\ \hline
        & 9 & Obtain informed consent and informing about the data use to the data subjects. & S* & Clearly document the process for obtaining informed consent from patients providing skin images or other relevant data. Ensure that participants understand the purpose of the dataset, how their data will be used, and their rights regarding data withdrawal.  \\ \hline
        & 10 & Do an analysis of the potential impact of the dataset and its use in data subjects. & S* & S* \\ \hline
 \end{tabular} %
    }
    \caption{PRISMA-Equity: Recommendations for curating a diverse dataset - Data Collection. *S - The same recommendation is applicable in this context as the standard.}
    \label{tab:dataset-recommendations-2}
    \vspace{-4mm}
\end{table}

\begin{table}[H]
    \centering
    \resizebox{\columnwidth}{!}{%
     \begin{tabular}{p{0.15\linewidth} | p{0.06\linewidth}|p{0.3\linewidth} |p{0.3\linewidth} |p{0.3\linewidth} }
        \textbf{Category} & \textbf{Item} & \textbf{Standard Image Curation with Ethical Focus} & \textbf{Extension for Healthcare} & \textbf{Extension for Melanoma Detection} \\ \hline
        \textbf{Data Annotation} & & & & \\ 
        \hline 
        & 11 & Provide guidelines for accurate data annotation. Describe how data will be associated with each instance. Describe methods and procedures to acquire the data. Define who will be involved in the data collection process. Specify the timeframe for collecting the data. & Involve healthcare experts from diverse backgrounds. & Involve expert dermatologists from diverse backgrounds. \\ \hline
        & 12 & Include demographic markers on the dataset. & S* & Include skin tone and skin hue as a marker in the dataset. Consider skin tone scales that consider lighter and darker skin tones. E.g. MONK skin tone scale. \\ \hline
         \textbf{Data Preprocessing and Analysis} & & & & \\ \hline
        & 13 & Describe steps taken for preprocessing the data. & S* & S*  \\ \hline
        & 14 & Indicate if duplicate images were removed. & S* & S*  \\ \hline
        & 15 & Indicate if anonymization and pseudonymization techniques were applied to the data to protect individual identities. & S* & S*  \\ \hline
        & 16 & Consider including recommendations for data split. & S* & S* \\ \hline
  \end{tabular} %
    }
    \caption{PRISMA-Equity: Recommendations for curating a diverse dataset - Data Annotation, Pre-processing, and Analysis. *S - The same recommendation is applicable in this context as the standard.}
    \label{tab:dataset-recommendations-3}
    \vspace{-4mm}
\end{table}

\begin{table}[H]
    \centering
    \resizebox{\columnwidth}{!}{%
     \begin{tabular}{p{0.15\linewidth} | p{0.06\linewidth}|p{0.3\linewidth} |p{0.3\linewidth} |p{0.3\linewidth} }
    \textbf{Category} & \textbf{Item} & \textbf{Standard Image Curation with Ethical Focus} & \textbf{Extension for Healthcare} & \textbf{Extension for Melanoma Detection} \\ \hline
    
    \textbf{Ideation stage} & & & &  \\ \hline
    
    & 1 & Define the purpose and intended use of the model. & S* & S*  \\ \hline
    
    & 2 & Engage with potential users and domain experts. & S* & Include dermatologists, oncologists, patients, healthcare providers and medical ethicists. \\ \hline
    
    & 3 & Consider risks and opportunities presented by an AI system. Examples opportunities: cultural preservation, enhanced public safety. Examples risks: Deepfakes and misinformation, consent and data misuse. & Consider risks and opportunities presented by an AI system in healthcare. Examples opportunities: improved healthcare diagnosis, accessibility. Examples risks: privacy invasions, bias and discrimination, misinterpretation and errors. & Consider risks and opportunities presented by an AI system to detect melanoma. Examples for opportunities: Early detection and improved accuracy. Consistency in diagnosis - Examples for risks: Bias and fairness when system is trained predominantly on data from certain demographics. Accuracy and reliability concerns when model is not trained on diverse and comprehensive dataset. Over-reliance on technology. \\ \hline
    
    & 4 & Identify and address potential biases that could influence the research direction like anchoring, availability heuristic, funding bias, group think, etc. & S* & Be cautious of biases that may arise from initial findings or prior research on melanoma detection. Regularly review and update methodologies based on new research and diverse data sources to ensure objectivity.  \\ \hline
    
    & 5 & Ensure representation of diverse demographic groups. & Ensure representation of diverse demographic groups. & Ensure representation of diverse demographic groups, including those with different skin types, to account for variations of melanoma appearance. \\ \hline
    \textbf{Data preparation} &  &  &  &  \\ \hline

    & 6 & Identify the groups or classes that need to be protected from biases and any proxy. Consider removing and saving any direct demographic markers and save them only for bias testing. & S* - For some medical systems or treatments, it might be important to keep those classes in the model. & S* \\ \hline
    
    & 7 & Test for representativeness and unbiased distribution of outcomes across demographic groups. & S* & S*  \\ \hline
    
    & 8 & Rebalance or reweigh the data as needed, ensuring legal and ethical compliance. & S* & S*  \\ \hline

    \end{tabular} %
    }
    \caption{PRISMA-Equity: Recommendations for inclusive model development - Ideation and Data Preparation Stages. *S - The same recommendation is applicable in this context as the standard.}
    \label{tab:model-recommendation}
    \vspace{-4mm}
\end{table}

\begin{table}[H]
    \centering
    \resizebox{\columnwidth}{!}{%
    \begin{tabular}{p{0.15\linewidth} | p{0.06\linewidth}|p{0.3\linewidth} |p{0.3\linewidth} |p{0.3\linewidth} }
    \textbf{Category} & \textbf{Item} & \textbf{Standard Image Curation with Ethical Focus} & \textbf{Extension for Healthcare} & \textbf{Extension for Melanoma Detection} \\ \hline
    \textbf{Model training and validation} & & & & \\ \hline
    
    & 9 & Consider robust techniques for validating your model like k-fold cross validation to assess model performance on different subsets of the data. & S* & S*  \\ \hline
    
    & 10 & Utilize appropriate metrics to evaluate model performance. & Utilize appropriate metrics to evaluate model performance like sensitivity, specificity, and area under the ROC curve (AUC). & S* \\ \hline
    
    & 11 & Based on the context/purpose of your model, its outcome choose the most adequate fairness metric to run bias testing. & S* - Metrics like equality opportunity and equalized odds, demographic parity, disparate impact, treatment equality can be relevant for healthcare AI models. & S* - Metrics like equal opportunity or equalized odds can help evaluate the model performance across skin tones. Testing for demographic parity, predicted value parity or treatment equality metrics could also be relevant. \\ \hline
    \textbf{Model interpretability} & & & & \\ \hline
    
    & 12 & Employ techniques like adversarial models and explainable AI to identify and understand any biases. For example model agnostic techniques like Shap values, LIME and feature importance visualizations like saliency maps or Grd-CAM. & S* & S*  \\ \hline
    
    \textbf{Model testing} & & & & \\ \hline
    
    & 13 & Consider holdouts and external tests and analyze the model performance on those datasets before making your final conclusions. & S* & S* - or releasing the model as a diagnosis tool. \\

    \end{tabular} %
    }
    \caption{PRISMA-Equity: Recommendations for inclusive model development - Model training, interpretability, and testing stages. *S - The same recommendation is applicable in this context as the standard.}
    \label{tab:model-recommendation-3}
    \vspace{-4mm}
\end{table}

\begin{table}[H]
    \centering
    \resizebox{\columnwidth}{!}{%
     \begin{tabular}{p{0.15\linewidth} | p{0.06\linewidth}|p{0.3\linewidth} |p{0.3\linewidth} |p{0.3\linewidth} } 
    \textbf{Category} & \textbf{Item} & \textbf{Standard Image Curation with Ethical Focus} & \textbf{Extension for Healthcare} & \textbf{Extension for Melanoma Detection} \\ \hline
    
    \textbf{Ethical considerations} & & & & \\
    
    & 14 & Document any ethical consideration considered in model development. Eg. use of sensitive data, mitigation strategies used, etc. & S* & Document efforts to identify and mitigate bias in melanoma detection models. Explain how diverse datasets are used to train and validate models, ensuring representation of various skin types, melanoma stages, and demographic groups. Describe any fairness assessments conducted and adjustments made to improve model equity.  \\ \hline
    
    \textbf{Model reporting} & & & & \\ \hline
    
    & 15 & Gather model details and document them. Consider the details on Model Cards for Model Reporting as key information to include in your report. Report any caveats and recommendations. & Provide a detailed description of the model architecture, including the type of model (e.g., deep learning, ensemble methods), key parameters, and any modifications made for healthcare applications. Report on the model’s performance metrics such as accuracy, precision, recall, and F1-score. Include metrics relevant to healthcare settings, such as clinical sensitivity and specificity. Clearly outline the intended use case for the model in healthcare, any limitations observed, and potential impacts on clinical decision-making. Provide recommendations for users regarding the model’s implementation, including best practices for deployment, potential adjustments needed for specific healthcare contexts, and any ongoing monitoring or validation required. & Detail the specific model used for melanoma detection, including the type (e.g., convolutional neural network, transfer learning model), key parameters, and any customizations for analyzing skin images and detecting melanoma. Include relevant performance metrics such as accuracy, sensitivity, specificity, and ROC-AUC scores. Highlight metrics related to melanoma detection, such as the model’s ability to differentiate between malignant and benign lesions. Describe the intended application of the model in melanoma detection, such as screening or diagnostic support. Note any limitations, such as performance variability across different skin types or lesion types. Document any caveats regarding the model’s performance, including limitations in detecting rare melanoma types or challenges in generalizing across diverse populations. Offer recommendations for effective use of the model, including guidance on integrating it into clinical workflows, considerations for retraining or updating the model, and monitoring for performance drift over time. \\ \hline
    \end{tabular} %
    }
    \caption{PRISMA-Equity: Recommendations for inclusive model development - Model Reporting and Ethical Considerations. *S - The same recommendation is applicable in this context as the standard.}
    \label{tab:model-recommendation-5}
    \vspace{-4mm}
\end{table}

\bibliographystyle{unsrt}  
\bibliography{ai-melanoma}

\end{document}